\documentclass[a4paper,12pt]{article}
\usepackage{bm}
\usepackage{latexsym}
\usepackage{dcolumn}
\usepackage{amsmath,amsfonts,amssymb}
\usepackage{graphicx,epsfig}
\usepackage{color}
\usepackage[active]{srcltx}%DO NOT DELETE

\def\l{\left}
\def\r{\right}
\def\DM{\mathrm{d}}

\def\eq#1{{Eq.~(\ref{#1})}}

\newcommand{\LL}{Lanczos-Lovelock }

\newcommand{\mdof}{microscopic degrees of freedom}

\newcommand{\bh}{black hole}

\usepackage{bm}
\usepackage{latexsym}
\usepackage{dcolumn}
\usepackage{amsmath,amsfonts,amssymb}
\usepackage{graphicx,epsfig}
\usepackage{color}
\def\eq#1{{Eq.~(\ref{#1})}}

\def\beq{\begin{equation}}
\def\eeq{\end{equation}}
\def\bea{\begin{eqnarray}}
\def\eea{\end{eqnarray}}
\def\benu{\begin{enumerate}}
\def\eenu{\end{enumerate}}
\def\nn{\nonumber}

\def\l{\left}
\def\r{\right}

\reversemarginpar

\def\DM{\mathrm{d}}
\begin{document}
%%%%%%%%%%%%%%%%%%%%%%%%%%%%%%%%%%%%%%%%%%%%%%%%%%%%%%%%%%%%%%%%%%%%%%%%%%%%%%%%%%%%%%%%%%%%%%%%%%%%%%%%
\title{Two Aspects of Black hole entropy in \LL models of gravity}
%%%%%%%%%%%%%%%%%%%%%%%%%%%%%%%%%%%%%%%%%%%%%%%%%%%%%%%%%%%%%%%%%%%%%%%%%%%%%%%%%%%%%%%%%%%%%%%%%%%%%%%%
\author{Sanved Kolekar$^{\ a,}$\footnote{sanved@iucaa.ernet.in}~~,~ Dawood Kothawala$^{\ b,}$\footnote{dawood.ak@gmail.com}~ and T.~Padmanabhan$^{\ a,}$\footnote{paddy@iucaa.ernet.in}\\ \\ \\
$^{a }$\ IUCAA, Pune University Campus, Ganeshkhind,\\
Pune 411007, India. \\
$^{b }$ \ Department of Mathematics and Statistics,\\
University of New Brunswick,\\
Fredericton, NB, Canada E3B 5A3
}

\date{\today}
\maketitle
%%%%%%%%%%%%%%%%%%%%%%%%%%%%%%%%%%%%%%%%%%%%%%%%%%%%%%%%%%%%%%%%%%%%%%%%%%%%%%%%%%%%%%%%%%%%%%%%%%%%%%%%
\begin{abstract}
%%%%%%%%%%%%%%%%%%%%%%%%%%%%%%%%%%%%%%%%%%%%%%%%%%%%%%%%%%%%%%%%%%%%%%%%%%%%%%%%%%%%%%%%%%%%%%%%%%%%%%%%

We consider two specific approaches to evaluate the black hole entropy which are known to produce correct results in the case of Einstein's theory and generalize them to \LL\ models. In the first approach (which could be called extrinsic)
 we use a procedure motivated by earlier work by Pretorius, Vollick and Israel, and by Oppenheim, and evaluate the entropy of a configuration of densely packed gravitating shells on the verge of forming a \bh\ in \LL theories of gravity. We find that this matter entropy is not equal to (it is less than) Wald entropy, except in the case of Einstein theory, where they are equal. The matter entropy is  proportional to the Wald entropy if we consider a 
specific $m$-th order \LL\ model,
with the proportionality constant depending on  the spacetime dimensions $D$ and the order $m$ of the \LL theory as $(D-2m)/(D-2)$. 
Since the proportionality constant depends on $m$, the proportionality between matter entropy and Wald entropy breaks down when we consider a sum of \LL\ actions involving different $m$.

In the  second approach (which could be called intrinsic) we generalize a procedure, previously introduced by Padmanabhan in the context of GR, to study off-shell entropy of a class of metrics with horizon using a path integral method.
We consider the Euclidean action of \LL\ models for a class of metrics off-shell and interpret it as a partition function. We show that in the case of spherically symmetric metrics, one can interpret the Euclidean action as the free energy and read off both the entropy and energy of a black hole spacetime. Surprisingly enough, this leads to exactly the Wald entropy and the energy of the spacetime in \LL\ models obtained by other methods. We comment on possible implications of the result.

\end{abstract}
%%%%%%%%%%%%%%%%%%%%%%%%%%%%%%%%%%%%%%%%%%%%%%%%%%%%%%%%%%%%%%%%%%%%%%%%%%%%%%%%%%%%%%%%%%%%%%%%%%%%%%%%
\section{Introduction}
%%%%%%%%%%%%%%%%%%%%%%%%%%%%%%%%%%%%%%%%%%%%%%%%%%%%%%%%%%%%%%%%%%%%%%%%%%%%%%%%%%%%%%%%%%%%%%%%%%%%%%%%
The first indication of the connection between thermodynamics and gravity came with the work of Bekenstein
 \cite{bekentropy} who proposed the idea that a black hole should have an entropy that is proportional to the area of its horizon. 
This interpretation became well established with the discovery of the temperature of the black hole \cite{hawking}.
Work in the last several decades attempted to understand the physical origin of the thermodynamic variables attributed to the horizons concentrating mostly on black hole horizons. In spite of extensive work and different possible suggestions for the source of, for example, entropy it is probably fair to say that we still do not quite understand the physics behind this phenomenon.

It is probably useful to classify the different approaches to explain \bh\ entropy by separating them into two broad categories (i) extrinsic origin - from the entropy of matter forming the \bh, entropy of matter fields propagating in the background metric, etc and (ii) intrinsic origin - microscopic degrees of freedom corresponding to underlying statistical theory of quantum gravity which are different depending on the approach. Within the context of Einstein's gravity, it is very difficult to discriminate between these two approaches. This is because, in Einstein's theory, entropy of a horizon is proportional to its area which has a simple geometrical meaning. It is therefore very easy to come up with completely different approaches (intrinsic as well as several extrinsic approaches) all of which will lead to $S\propto A$. 

On the other hand, the proportionality between horizon entropy and area does not extend to more general class of gravitational theories in which the entropy is given by a prescription due to  Wald \cite{Wald} which essentially identifies the horizon entropy with a suitably defined Noether charge. 
Many of the approaches which correctly reproduces $S\propto A$ in the context of GR cannot be generalized in a natural fashion to more general class of theories like, for example, \LL\ models. (One such  example, which does \textit{not} generalize, is the entanglement entropy; see, for example, Ref. \cite{TPentangle}. Also see \cite{naresh} which discusses the so-called \textit{universality} of entropy in pure Lovelock gravity.)
Therefore, the possibility of generalization beyond GR acts as an acid test in discriminating between the different approaches for obtaining the horizon entropy both intrinsic and extrinsic. 

In the limited context of understanding black hole entropy, one may think of these various possibilities as just theoretical curiosity and may take the point of view that any one, valid, derivation is good enough. But the entire topic has assumed a far greater significance in recent years with 
several  results strongly indicating the possibility that the field equations of gravity have the same status as the equations of fluid mechanics or elasticity. (For a recent review, see Ref. \cite{paddyaspects}.)
This approach has a long history originating from the work of Sakharov \cite{sakharov} and interpreted in many ways by different authors (for a incomplete sample of references, see Ref. \cite{others}). One specific implementation of this idea considers the \textit{field equations} of the theory to be  `emergent' in a well-defined sense, rather than use that term in a more speculative vein --- like e.g., considering the space and time themselves to be emergent etc.
The evidence for such a specific interpretation comes from different facts like the possibility of interpreting the field equation in a wide class of theories as thermodynamic relations \cite{paddy2002,LLeom}, the nature of action functional in gravitational theories and their thermodynamic interpretation \cite{holoaction}, the possibility of obtaining the field equations from a thermodynamic extremum principle \cite{aseemtp}, application of equipartition ideas to obtain the density of microscopic degrees of freedom \cite{equiv}, the equivalence of Einstein's field equations to the Navier-Stokes equations near a null surface \cite{NS,navieractionsanved} etc. 
In such an approach, one works with local Rindler horizons as analogues of on-shell horizons like black hole horizons and attempts to introduce the concept of entropy density in spacetime. The precise nature of horizon entropy (whether it is intrinsic to spacetime or related to matter degrees of freedom) assumes far greater significance in this context.

In  this paper, we have studied two approaches to \bh\ entropy, one intrinsic and the other extrinsic, in the general context of \LL\ models of gravity. Both these approaches have been previously shown to reproduce the correct horizon entropy in Einstein's theory. When generalized to \LL\ models the extrinsic approach fails to give the correct Wald entropy of the theory (though, for a pure $m$-th order \LL model it is a near miss with the entropy being proportional to Wald entropy)
while the intrinsic approach gives exactly the correct result for not only the entropy but even the energy associated with the black hole in \LL models!

We shall now briefly describe the two routes to the entropy which we consider in this paper.
The most natural extrinsic origin of \bh\ entropy would be to consider the entropy of the matter which formed the \bh. Indeed, such an approach was taken in \cite{areascalingisrael, areascalingoppenheim} where it was shown that when a system consisting of spherically symmetric gravitating shell, or a series of such shells forming a star, is on the verge of forming a \bh, then its entropy is proportional to the area of the outermost shell which is same as the area of the horizon it would form at the end of its collapse. Further the proportionality constant was calculated to be $1/4$ making the entropy of the star to be exactly equal to the Bekenstein-Hawking entropy of the \bh. (More recent work \cite{areaentropysanved} suggests that there could be a purely kinematic reason for some of these results.) In the first part of this paper, we extend this work to \LL\ models of gravity and 
calculate the entropy of a spherically symmetric gravitating star on the verge of forming a \bh, in \LL\ gravity. (We also compute the same for 
 $F(R)$ theories; the result is given in Appendix \ref{app:fR}.) We find that the entropy of the system is, in general, \textit{not} equal to the Wald entropy. However, at each order of the \LL\ model, labeled by an index $m$, with $m=1$ being Einstein's theory, $m=2$ being Gauss-Bonnet etc.,  the entropy which we obtain is
proportional to Wald entropy of the \bh\ horizon in the corresponding theory of gravity with the proportionality constant 
depending on the dimension of spacetime as well as $m$. So when we consider a \LL\ theory with different orders like, for example, Einstein plus Gauss-Bonnet kind of theories, the overall proportionality breaks down.  That is, for a \LL theory described by a sum of, say, first $m$ \LL terms, the matter entropy of the shell configuration will \textit{not} even be proportional to Wald entropy. (Unlike \LL\ models, the $F(R)$ models lead to an equality between the entropy computed by this method and Wald entropy.)

In part 2 of the paper, 
we  turn our attention to an ``intrinsic" approach to horizon entropy, which attempts to interpret the \bh\ entropy as associated with statistical nature of \mdof\ of some underlying statistical theory of gravity. One such approach was considered in \cite{paddy2002}, where a partition function was defined through a Euclideanization of the path integral for Einstein's gravity. Here, the Euclidean action was interpreted to be the effective action for gravity emerging from some unknown quantum theory. It was shown that the form of the partition function allows one to determine the entropy \textit{and} energy associated with the horizon which is same as determined through other approaches. However, the crucial difference between this approach  and others is that here one does not use the field equations of gravity in the derivation thus allowing us to define an entropy and energy for the horizon even off-shell. In contrast, all the other approaches to entropy rely mainly on the field equations providing us with only a on-shell definition of entropy. In the second part of this paper, in section 3, we consider 
the generalization of this ``intrinsic'' approach to black hole entropy to \LL\ models. Surprisingly enough, we find that the result generalizes in a nice manner to all the \LL\ models. We find that the entropy and energy associated with the horizon off-shell are exactly same as the Wald entropy and energy of a horizon as obtained through other approaches on-shell.

We shall work in $D$ spacetime dimensions. Latin indices $a, b, \ldots = 0$ to $(D-1)$, Greek indices $\mu, \nu, \ldots = 0, 2, 3, \cdots (D-1)$, and capitalized Latin indices $A, B, \ldots = 2$ to $(D-1)$.

%%%%%%%%%%%%%%%%%%%%%%%%%%%%%%%%%%%%%%%%%%%%%%%%%%%%%%%%%%%%%%%%%%%%%%%%%%%%%%%%%%%%%%%%%%%%%%%%%%%%%%%%
\section{Entropy of gravitating system}
%%%%%%%%%%%%%%%%%%%%%%%%%%%%%%%%%%%%%%%%%%%%%%%%%%%%%%%%%%%%%%%%%%%%%%%%%%%%%%%%%%%%%%%%%%%%%%%%%%%%%%%%

To calculate the standard thermodynamic entropy of a self gravitating system, we follow the set up suggested by Oppenheim \cite{areascalingoppenheim} motivated by earlier work of Israel et al \cite{areascalingisrael}. (A different setup leading to the same result in Einstein gravity is discussed in \cite{lemos}.) We consider a system of $n$ densely packed spherically symmetric shells in $D$ dimensions, assumed to be in thermal equilibrium, and supporting itself against its own gravity. We shall be interested in the entropy of this system when the outermost shell is close to the event horizon of the system.

%%%%%%%%%%%%%%%%%%%%%%%%%%%%%%%%%%%%%%%%%%%%%%%%%%%%%%%%%%%%%%%%%%%%%%%%%%%%%%%%%%%%%%%%%%%%%%%%%%%%%%%%
\subsection{The set-up}
%%%%%%%%%%%%%%%%%%%%%%%%%%%%%%%%%%%%%%%%%%%%%%%%%%%%%%%%%%%%%%%%%%%%%%%%%%%%%%%%%%%%%%%%%%%%%%%%%%%%%%%%

We first describe a general set-up to study the system described above for any spherically symmetric spacetime. This is important for two reasons. First, as we shall see, it will highlight the key mathematical feature responsible for the entropy density $S/A$ of the system being $1/4$ for Einstein's theory in arbitrary dimensions. In \cite{areascalingoppenheim}, this result was obtained for $D=4$ by explicitly using the Schwarzschild form of metric functions and the expression for Hawking temperature. However, we shall not require any explicit expression for metric functions or Hawking temperature to obtain the area scaling. Second, the set-up we describe can be used, without any modification, to obtain entropy density for higher curvature gravity actions.

Let us denote the variables or parameters describing the $i$th shell with, say, $X_i$. Since the system considered is spherically symmetric, we can write the metric outside the $i$th shell as

\begin{equation}
 ds^2 = -c_i f_i(r)dt^2 + b_i(r)^{-1}dr^2 + r^2 d\Omega^2
\label{shellmetric}
\end{equation}
where $d\Omega^2$ is the metric of a unit $(D-2)$ sphere. At the location of the shells (given by $r_i=$const), the metric above must satisfy the first Israel junction condition which states that the induced metric on the hypersurface should be continuous. 
This leads to the following conditions on the constants $c_i$:
\begin{equation}
 c_i f_i(r_{i+1}) = c_{i+1} f_{i+1}(r_{i+1})
\label{1junction}
\end{equation}
and we shall choose $c_n =1$ (which fixes the interpretation of $t$ as proper time for asymptotic observers). Then \eq{1junction} can be solved to give
\begin{eqnarray}
 c_k = \frac{f_{k+1}(r_{k+1})}{f_{k}(r_{k+1})}.....\frac{f_{n-1}(r_{n-1})}{f_{n-2}(r_{n-1})}\frac{f_{n}(r_{n})}{f_{n-1}(r_{n})}
\label{ccondition}
\end{eqnarray}
Note that when $f_{n}(r_{n}) = 0$ which is true when the outermost shell is exactly on the horizon we have $c_n=0$ for all $(i \neq n)$ which implies that 
\begin{equation}
 g_{00} = 0 \; \; \forall \ i
\label{g00vanish}
\end{equation}
Here the condition that the $g_{00}$'s vanish even for the inner-shells indicates that our assumption regarding the staticity of the inner shells is not valid when $f_{n}(r_{n})$ is exactly zero since we know that a particle cannot be kept at a fixed position inside a blackhole without letting it fall into the singularity. However for the purpose of our discussion, we only need to consider the limit in which the outer shell is very near to the horizon, that is, $f_{n}(r_{n}) \rightarrow 0$, then the statement of $c_i$'s (other than $c_n$) and $g_{00}$ being equal to zero is just the leading order term in this approximation. Henceforth, we shall assume that we are working in this limit and will not state it explicitly unless otherwise needed.

We shall next use the second junction condition which relates jumps in geometric quantities across a shell, to the matter stress tensor of the shell, $t_{(i)\mu \nu}$. Denote the normal to a $r=$const. surface (dropping the subscript $i$ for convenience) by $n_{a} = \l(1/\sqrt{b}\r) \partial_a r$, so that the induced metric becomes $h_{\mu \nu} = \l( g_{a b} - n_a n_b \r) \delta^a_{\mu} \delta^b_{\nu}$ in coordinates $(t, \theta_A)$. Within the hypersurface, we can further define $u_{\mu} = \sqrt{c f} \; \partial_{\mu} t$, and the induced metric on the level surfaces of $t$, $q_{\mu \nu} = h_{\mu \nu} + u_{\mu} u_{\nu}$. It then follows from spherical symmetry that $t_{\mu \nu}$ for each shell has the general form
\begin{eqnarray}
t_{(i)\mu \nu} = \rho_i u_{(i)\mu} u_{(i)\nu} + P_i q_{(i)\mu \nu}
\label{eq:gensab}
\end{eqnarray}
with $E_i = 4\pi r_i^2 \rho_i$ is the energy of the shell. The physical interpretation of $E_i$ and $P_i$ is that of energy and pressure as measured by a local observer at rest on the shell. Further, the condition for thermal equilibrium in curved spacetime implies \cite{tolman}
\begin{eqnarray}
 T_i \sqrt{-g^i_{00}(r_i)} &=& T_{\infty}
 \nn \\
\mu_i \sqrt{-g^i_{00}(r_i)} &=& \mu_n \sqrt{-g^n_{00}(r_n)}
\label{tolman-chemical}
\end{eqnarray}
where $T_{\infty}$ is the temperature of the system as measured by a static observer at infinity, and $\mu_i$'s denote the chemical potential. We can thus express all the $T_i$'s and $\mu_i$'s in terms of just two unknown parameters $T_{\infty}$ and $\mu_n$ respectively. Further, in thermodynamic equilibrium, each shell satisfies the Gibb's Duhem relation
\begin{equation}
 E_i = T_i S_i - P_i A_i + \mu_i N_i
\end{equation}
where $N_i$ is the number of particles composing the $i$th shell and $A_i$ is the hypersurface area of the $i$th shell. Using Eqs.~(\ref{tolman-chemical}) in the above expression, we can write the total entropy $S_{\mathrm{matter}}$ of the system in the form
\begin{eqnarray}
 S_{\mathrm{matter}} &=& \sum_i S_i
 \nn \\
 &=& \sum_i \frac{E_i + P_i A_i}{T_{\infty}}\sqrt{-g^i_{00}(r_i)} - \frac{\mu_n N}{T_{\infty}}\sqrt{-g^n_{00}(r_n)}
 \nn \\
  &=& \sum_i \frac{E_i + P_i A_i}{T_{\infty}}\sqrt{c_i f_i(r_i)} - \frac{\mu_n N}{T_{\infty}}\sqrt{f_n(r_n)}
 \nn \\
\label{eq:Stotshell}
\end{eqnarray}
Now assuming that $\mu_n N$ is a finite quantity, the last term in the above expression vanishes in the near horizon limit, since $g^n_{00}(r_n)$ vanishes (see \eq{g00vanish}). Hence the only non-trivial contribution to $S_{matter}$ can come from the first term, which we wish to evaluate in the limit when outermost shell is close to the horizon of the system, $r_n \rightarrow r_H$ (where $f_n(r_H)=0$); we shall refer to this as the near-horizon limit. We can ``read-off" this contribution as follows. First, we note that $(E_i + P_i A_i)/T_{\infty}$ has no dependence on $c_i$, since $t_{(i)\mu \nu}$ does not depend on $c_i$ (see below). Further, in the near horizon limit, $c_i=0 ~ \forall i \neq n, c_n=1$, from which it is easy to see that the only non-zero contribution to $S_{\mathrm{matter}}$ will come from the $i=n$ term, and in particular from those terms in $(E_n + P_n A_n)$ which diverge as $1/\sqrt{f_n(r_n)}$. Therefore, our strategy to calculate the entropy of the system will be to analyze the form of $E_n$ and $P_n$ and look for such divergent factors. As we shall now demonstrate, this strategy provides an extremely quick way to obtain the result in Einstein theory (compared to the explicit computations in \cite{areascalingoppenheim}), and also facilitates easy generalization to higher derivative theories discussed in the next section.

In Einstein theory, the surface stress tensor is given by
\begin{eqnarray}
8 \pi t^{(E)}_{(i)\mu \nu} &=& \Biggl \langle K h_{\mu \nu} - K_{\mu \nu} \Biggl \rangle_i
\label{eq:sabeins}
\end{eqnarray}
where $\langle \ldots \rangle_i = \langle \ldots \rangle_{i+1} - \langle \ldots \rangle_i$ evaluated at $r=r_i$. The extrinsic curvature of $r=$const surface, for the metrics described by Eq.~(\ref{shellmetric})), is given by (after dropping the subscript $i$ for convenience)
\begin{eqnarray}
K_{\mu \nu} &=& - \l( \sqrt{b} f'/2 f \r) u_{\mu} u_{\nu} + \l(\sqrt{b}/r\r) q_{\mu \nu}
\nn \\
\nn \\
K &=& h^{\mu \nu} K_{\mu \nu}
\nn \\
&=& \l( \sqrt{b} f'/2 f \r) + \l(D-2\r) \l(\sqrt{b}/r\r)
\label{eq:extcurv}
\end{eqnarray}
Note that there is no dependence on $c_i$ as stated above. Based on our strategy outlined in the previous section, we must now find terms which, for $i=n$, diverge as $1/\sqrt{f_n(r_n)}$ as $r_n \rightarrow r_H$. We now specialize to the case $b_i(r)=f_i(r) ~ \forall i$. \footnote{Actually, this restriction is not necessary since, in the near-horizon limit, it is automatically satisfied for all the cases in which the horizon has no curvature singularity.} Using Eqs.~(\ref{eq:sabeins}), (\ref{eq:extcurv}) and (\ref{eq:Stotshell}), we can immediately write, for the divergent parts,
\begin{eqnarray}
8 \pi [ \rho_n]_{\mathrm{div}} = 8 \pi \l[ t^{(E)}_{(n)\hat 0 \hat 0} \r]_{\mathrm{div}} &=& 8 \pi \l[ t_{(n)\mu \nu} u^{\mu} u^{\nu} \r]_{\mathrm{div}} = 0 
\nn \\
\nn \\
8 \pi [ P_n]_{\mathrm{div}} \; \delta^A_{B} = 
8 \pi \l[ t^{(E)A}_{(n)B} \r]_{\mathrm{div}} &=& - K^{(n)}_{\hat 0 \hat 0} \delta^A_{B} 
\label{eq:sabeinsdiv}
\end{eqnarray}
in the limit $r_n \rightarrow r_H$, $f_n(r_H) \rightarrow 0$ (here $K^{(n)}_{\hat 0 \hat 0} = K^{(n)}_{\mu \nu} u^{\mu} u^{\nu} $). Note that we have recalled the definition of $\rho_i$'s and $P_i$'s from Eq. (\ref{eq:gensab}). Hence, we see that the main divergence comes from 
\begin{eqnarray}
\l[ P_n \r]_{\mathrm{div}} &=& - \l( \frac{1}{8 \pi} \r) K^{(n)}_{\hat 0 \hat 0} 
\nn \\
&=& \l( \frac{1}{16 \pi} \r) \lim_{r_n \rightarrow r_H^+} \frac{f_n'(r_n)}{\sqrt{f_n(r_n)}}
\nn \\
&=& \frac{T_{\infty}}{4}  \lim_{\epsilon \rightarrow 0^+} \frac{1}{ \sqrt{f_n(r_H + \epsilon)} }
\label{eq:PTeos}
\end{eqnarray}
where we have used the definition of Hawking-Unruh temperature for a system which has collapsed to a black hole. Using the above result in Eq. (\ref{eq:Stotshell}), we finally get
\begin{eqnarray}
S_{\mathrm{matter}} &\underset{r_n \rightarrow r_H}{\longrightarrow}& S_n
\nn \\
&=& \lim_{\epsilon \rightarrow 0^+} \frac{(1/4) T_{\infty} / \sqrt{f_n(r_H+\epsilon)}}{T_{\infty}} A_H \times \sqrt{f_n(r_H+\epsilon)}
\nn \\
&=& \frac{A_H}{4}
\end{eqnarray}
Hence, we see that the entropy of the configuration, as the outermost shell approaches the event horizon of the system, is dominated by the Bekenstein-Hawing entropy. The crucial factor of $1/4$ comes from the last line of Eq.~(\ref{eq:PTeos}), which can be recast as 
\begin{equation}
 [P_n]_{\mathrm{div}}/T_n = 1/4
\label{eqnofstateeg}
\end{equation}
Interestingly, this same relation was recently arrived at in the context of an action based derivation of Navier-Stokes equations and its relation to emergent gravity paradigm \cite{navieractionsanved}.

In what follows, we shall repeat the above analysis for two specific class of higher curvature gravity theories, namely \LL gravity and $F(R)$ models (which we put in Appendix \ref{app:fR} since it is only a slight variation of the calculation in Einstein gravity discussed above), and show that the correspondence between shell entropy as calculated above, and the black hole entropy as given by Wald formula (which is a generalization of Bekenstein-Hawking entropy to higher derivative gravity actions) {\it does not always hold}.

%%%%%%%%%%%%%%%%%%%%%%%%%%%%%%%%%%%%%%%%%%%%%%%%%%%%%%%%%%%%%%%%%%%%%%%%%%%%%%%%%%%%%%%%%%%%%%%%%%%%%%%%
\section{Higher derivative theories and comparison with Wald entropy}
%%%%%%%%%%%%%%%%%%%%%%%%%%%%%%%%%%%%%%%%%%%%%%%%%%%%%%%%%%%%%%%%%%%%%%%%%%%%%%%%%%%%%%%%%%%%%%%%%%%%%%%%

We now proceed to the \LL case taking exactly the same route as in Einstein case described above. We will also point out (in the final subsection) that our result has a curious interpretation in terms of ``equation of state" of the horizon degrees of freedom, and also a very direct mathematical connection with the membrane paradigm for black hole horizons (explored more fully in a separate publication \cite{membraneLL}).

%%%%%%%%%%%%%%%%%%%%%%%%%%%%%%%%%%%%%%%%%%%%%%%%%%%%%%%%%%%%%%%%%%%%%%%%%%%%%%%%%%%%%%%%%%%%%%%%%%%%%%%%
\subsection{\LL gravity}
%%%%%%%%%%%%%%%%%%%%%%%%%%%%%%%%%%%%%%%%%%%%%%%%%%%%%%%%%%%%%%%%%%%%%%%%%%%%%%%%%%%%%%%%%%%%%%%%%%%%%%%%

The $m^{\mathrm{th}}$ order \LL lagrangian $L_m$ is given by completely anti-symmetrised product of $m$ curvature tensors 
\begin{eqnarray}
L_m^{(D)} = \frac{1}{16 \pi} \frac{1}{2^m} \delta^{a_1 b_1 \ldots a_m b_m}_{c_1 d_1 \ldots c_m d_m} R^{c_1 d_1}_{~ a_1 b_1} \cdots R^{c_m d_m}_{~ a_m b_m}
\end{eqnarray}
For $m=1$, $L_m$ reduces to $(16 \pi)^{-1} R$, which is the Einstein-Hilbert lagrangian. The surface stress tensor in the $m^{\mathrm{th}}$ order \LL theory is given by \cite{surfacestressLL}
\begin{eqnarray}
8 \pi \left( t^{\; \; \ \ \nu}_{(m) \mu} \right)_i &=&  \frac{m!}{2^{m+1}}\,\alpha _{m} \Biggl \langle \sum_{s=0}^{m-1}C_{s(m)}\,\left( \pi
_{s(m)}\right) ^{\nu}_{\mu} \Biggl \rangle_i \nonumber \\
\left( \pi _{s(m)}\right) ^{\nu}_{\mu} &=& \delta _{\lbrack
\mu \mu_{1}\cdots \mu_{2m-1}]}^{[\nu \nu_{1}\cdots \nu_{2m-1}]}\,\hat{R}_{\nu_{1}\nu_{2}}^{\mu_{1}\mu_{2}}\cdots \hat{R}_{\nu_{2s-1}\nu_{2s}}^{\mu_{2s-1}\mu_{2s}}%
\,K_{\nu_{2s+1}}^{\mu_{2s+1}}\cdots K_{\nu_{2m-1}}^{\mu_{2m-1}} 
\label{junctionconditionLL}
\end{eqnarray}%
where the coefficients $C_{s(m)}$ are given by
\begin{equation}
C_{s(m)}=\sum_{q = s}^{m-1} \frac{4^{m-q} \; ^qC_s (-2)^{q-s}}{q!\left( 2m-2q-1\right) !!}
\label{coefficient}
\end{equation}
Here $\hat R_{\mu \rho \sigma \nu}$ etc. indicates that these quantities are to be evaluated for the induced metric $h_{\mu \nu}$. One can check that for $m=1$ and $m=2$, the above expression reduces to that of the surface stress tensor in Einstein's gravity (see \eq{eq:sabeins}) and Gauss-Bonnet Gravity \cite{junctionGB} (see \eq{junctionconditionGB} in the appendix) respectively. 
To calculate the matter entropy of the system of $n$ shells, we proceed in the manner similar as in the case of Einstein's gravity described in the previous section, that is, by finding $\l[ t^{\ \ \nu}_{(n) \mu} \r]_{\mathrm{div}}$. We note that, in the near horizon limit (and assuming that the intrinsic Riemann tensor of horizon surface is finite), the only divergence in the stress tensor will come from the $K^{\hat 0}_{\hat 0}$ component of the extrinsic curvature of the $r= r_n$th shell, which diverges as $(1/\sqrt{f_n})$ as can be seen from \eq{eq:extcurv}. The transverse components $K^A_B$ vanish as $(\sqrt{f_n}/r)$. Further, note that the determinant tensor has the property 
\begin{equation}
\delta^{0 \alpha_1 \alpha_2 \cdots \alpha_n}_{0 \beta_1 \beta_2 \cdots \beta_n} = \delta^{A_1 A_2 \cdots A_n}_{B_1 B_2 \cdots B_n} \times \l( \delta^{\alpha_1}_{A_1} \delta^{B_1}_{\beta_1} \ldots \delta^{\alpha_n}_{An} \delta^{B_n}_{\beta_n} \r)
\label{determinanttensor}
\end{equation}
That is, the presence of $0$ in each row of the determinant tensor forces all the other indices to take the values $2,3, \cdots (D-1)$. Thus, at the most, we can have only one $K^0_0$ present in any term of \eq{junctionconditionLL} which shows that the maximum possible divergence is $O(1/\sqrt{f_n})$ even though the surface stress tensor is a $(2m-1)$th degree polynomial in $K$. Further, from the structure of \eq{junctionconditionLL}, it is easy to see that only the $s=(m-1)$ term gives this divergent contribution while all the other terms being of $O(\sqrt{f})$ or higher vanish. Further, from \eq{determinanttensor}, it is again obvious that only the transverse component of the surface stress tensor will contribute. Thus we get
\begin{eqnarray}
8 \pi [ \rho_n]_{\mathrm{div}} = 8 \pi \l[ t^{}_{(n)\hat 0 \hat 0} \r]_{\mathrm{div}} &=& 8 \pi \l[ t_{(n)\mu \nu} u^{\mu} u^{\nu} \r]_{\mathrm{div}} = 0 \nn \\ \nn \\
8 \pi [ P_n]_{\mathrm{div}} \; \delta^A_B = 
8 \pi \l[ t^{A}_{(n)B} \r]_{\mathrm{div}} 
&=& \frac{m \alpha _{m}}{2^{m-1}} \delta ^{
A A_{1}\cdots A_{2m-2}}_{B B_{1}\cdots B_{2m-2}}\,\hat{\hat{R}}^{B_{1}B_{2}}_{A_{1}A_{2}}\cdots \hat {\hat{R}}^{B_{2m-3}B_{2m-2}}_{A_{2m-3}A_{2m-2}} \; K^{\hat 0}_{\hat 0}
\label{sabdivergenceLL}
\end{eqnarray}
where in the second line we have used $C_{m-1} = 4/(m-1)!$ from its definition in \eq{coefficient}, and replaced $\hat{R}^{A B}_{C D}$ with the intrinsic curvature $\hat{\hat{R}}^{A B}_{C D}$ defined completely in terms of the induced metric $q_{AB}$ of the constant $t$, constant $r$ surface. Such a replacement is valid since the corresponding extrinsic curvature vanishes. We can now obtain the pressure easily using the fact that $q_{AB}$ is maximally symmetric (see \eq{eq:gensab}), and hence
\begin{eqnarray}
 [ P_n]_{\mathrm{div}} = \frac{1}{(D-2)} {q^{AB}\l[ t_{(n)AB} \r]_{\mathrm{div}}} &=& \frac{2 m \alpha_{m}}{(D-2)} \frac{1}{16 \pi} \frac{1}{2^{m-1}} \delta ^{
A A_{1}\cdots A_{2m-2}}_{A B_{1}\cdots B_{2m-2}}\,\hat{\hat{R}}^{B_{1}B_{2}}_{A_{1}A_{2}}\cdots \hat {\hat{R}}^{B_{2m-3}B_{2m-2}}_{A_{2m-3}A_{2m-2}} \; K^{\hat 0}_{\hat 0} \nonumber \\
&=& \l( \frac{D-2m}{D-2} \r) 2 m \alpha_{m} L^{(D-2)}_{m-1} \; K^{\hat 0}_{\hat 0}
\label{pressureLL}
\end{eqnarray}
In arriving at the second equality, we have used the following two relations concerning \LL actions of order $m$ in $D$ dimensions \cite{LLeom}
\begin{eqnarray}
{\mathcal G}^i_{j(m)} &=&  - \frac{1}{2} \frac{1}{16 \pi} \frac{1}{2^m} \delta^{i a_1 b_1 \ldots a_m b_m}_{j c_1 d_1 \ldots c_m d_m} R^{c_1 d_1}_{~ a_1 b_1} \cdots R^{c_m d_m}_{~ a_m b_m}
\nn \\
{\mathcal G}^i_{i(m)} &=& - \frac{D-2m}{2} L^D_m
\end{eqnarray}
where $G^i_{j(m)}$ is the equation of motion tensor for \LL action; for e.g., for $m=1$, ${\mathcal G}^i_{j(1)} = (16 \pi)^{-1} G^i_j$ where $G^i_j$ is the Einstein tensor. Hence, once again we find that the only non-zero contribution to matter entropy comes from the pressure term as in the case of Einstein gravity (see \eq{eq:sabeinsdiv}). No further computation needs to be done, since we can simply obtain the entropy by comparing expression (\ref{pressureLL}) with the first of Eqs.~(\ref{eq:PTeos}) in Einstein theory. This immediately allows us to ``read-off" the matter entropy as
\begin{eqnarray}
S^{(m)}_{\mathrm{matter}} = \l( \frac{D-2m}{D-2} \r) 4 \pi m \alpha_m L^{(D-2)}_{m-1} A_H \label{matterentropyLL}
\end{eqnarray}
where $A_H$ is the $D-2$ dimensional hypersurface area of the horizon. On the other hand, the Wald entropy for general \LL theory can be shown to give
\begin{eqnarray}
S^{(m)}_{\mathrm{Wald}} &=&  4 \pi m \alpha_m L^{(D-2)}_{m-1} A_H
\end{eqnarray}
which leads to the following relation
\begin{eqnarray}
 S^{(m)}_{\mathrm{matter}} = \l( \frac{D-2m}{D-2} \r) S^{(m)}_{\mathrm{Wald}}
\label{entropymatterwald}
\end{eqnarray}
Thus we find  that the matter entropy in \LL theory is proportional to the corresponding Wald entropy of the \bh. We also see that $S_{\mathrm{matter}} \leq S_{\mathrm{Wald}}$ with the equality holding only in Einstein's gravity $m=1$. This inequality strongly suggests that the \mdof \ responsible for the entropy of the \bh\ are certainly not the extrinsic (matter) degrees of freedom forming the \bh. Further, since a general \LL theory will be described by lagrangian of the form $L=\sum_m \alpha_m L_m$, the total matter entropy in such theories would be 
\begin{eqnarray}
 S_{\mathrm{matter}} = \sum \limits_{m=0}^{[D-1)/2]} \l( \frac{D-2m}{D-2} \r) S^{(m)}_{\mathrm{Wald}}
\end{eqnarray}
which has no simple relation, even proportionality to the Wald entropy 
\begin{eqnarray}
 S_{\mathrm{Wald}} = \sum \limits_{m=0}^{[D-1)/2]} S^{(m)}_{\mathrm{Wald}}
\end{eqnarray}

\subsection{Equation of state}
As pointed out above in the case of Einstein gravity as well as in the \LL models, it is only the pressure term which leads to a non-zero contribution to the matter entropy and hence, in the near horizon limit, we find that the following relation holds
\begin{eqnarray}
\frac{S_{\mathrm{matter}}}{A_H} &=& \frac{[P_n]_{\mathrm{div}}}{T_n} \nonumber \\
&\equiv& \frac{P}{T_n}
\end{eqnarray}
where for the sake of brevity, we have replaced $[P_n]_{\mathrm{div}}$ by $P$. Thus the scaling property of entropy with the horizon radius $r_H$, as well as the proportionality constant, is completely determined by the relation between $P$ and $T$, which is to say that the equation of state $P = P(T)$ determines the matter entropy. In the case of \LL gravity, we can read off the equation of state $P = P(T) = [t^{\ \ \theta}_{(n) \theta}]_{\mathrm{div}}$ from \eq{pressureLL} by using the relation $K_{\hat 0 \hat 0} = -K^{\hat 0}_{\hat 0} = -f^{\prime}_H/(2\sqrt{f}) = -2\pi T_n$ to get
\begin{eqnarray}
 \frac{P}{T_n} = \l( \frac{D-2m}{D-2} \r) 4 \pi m \alpha_m L^{(D-2)}_{m-1}
\label{generaleqnofstate}
\end{eqnarray}
Using \eq{entropymatterwald}, the above relation can also be written in terms of Wald entropy
\begin{eqnarray}
 \frac{P A_H}{T_n} = \l( \frac{D-2m}{D-2} \r) S^{(m)}_{\mathrm{Wald}}
\end{eqnarray}
Further, it is also interesting to note that the pressure that one obtains in the context of the membrane paradigm for both Einstein and Gauss Bonnet black holes is exactly the same as given by \eq{generaleqnofstate} for $m=1$ and $m=2$ respectively \cite{membranepressure}. In fact, we have presented an investigation of the full membrane paradigm for \LL\ theories in a separate publication \cite{membraneLL}, obtaining not only the membrane pressure but also other transport coefficients from the near-horizon surface stress tensor.

%%%%%%%%%%%%%%%%%%%%%%%%%%%%%%%%%%%%%%%%%%%%%%%%%%%%%%%%%%%%%%%%%%%%%%%%%%%%%%%%%%%%%%%%%%%%%%%%%%%%%%%%
\section{The partition function for \LL gravity}
%%%%%%%%%%%%%%%%%%%%%%%%%%%%%%%%%%%%%%%%%%%%%%%%%%%%%%%%%%%%%%%%%%%%%%%%%%%%%%%%%%%%%%%%%%%%%%%%%%%%%%%%
All the analysis in the literature regarding the definitions of $S$ and $E$ of a black hole uses the field equations of the corresponding theory of gravity. In our analysis in preceding section, although the bulk field equations were not used, the relation between geometric quantities describing the surface and the matter stress tensor on the surface can be viewed as equations of motion for the boundary variables. We will now discuss a second aspect of black hole entropy (and energy), in the context of spherically symmetric horizons in \LL theory, based on a quantum picture of gravity which does not involve the field equations of the theory. Such an approach was first suggested in \cite{paddy2002} in the case Einstein's theory, and our analysis is a generalization of the same to \LL theories. More specifically, we shall show that the Euclidean action $A^{\mathrm (e)}_{\mathrm{LL}}$ describing spherically symmetric \LL models gives the free energy and has the structure
\begin{eqnarray}
A^{\mathrm (e)}_{\mathrm{LL}} = \beta F = \beta E - S^{\mathrm (e)}
\end{eqnarray}
from which one can ``read-off" the entropy $S^{\mathrm (e)}$ and energy $E$. We shall show that these match exactly with Wald entropy (i.e., $S^{\mathrm (e)}=S_{\mathrm{Wald}}$) and energy derived by independent methods for these theories.

We again consider spherically symmetric spacetimes in $D$ dimensions
\begin{equation}
 ds^2 = -f(r)dt^2 + f(r)^{-1}dr^2 + r^2 d\Omega^2
\label{fmetric}
\end{equation}
such that $f(r)$ vanishes at some surface $r=a$, say, with $f'(a)$ ($= B$, say) remaining finite. Since the metric is static, Euclidean continuation is trivially affected by $t\to \tau=it$ and to avoid conical singularity at $r=a$, $\tau$ must be periodic with period $\beta=4\pi/|B|$, corresponding to the temperature $T=|B|/4\pi$. The class of metrics in \eq{fmetric} with the behaviour $[f(a)=0, f'(a)=B]$ constitute a canonical ensemble at constant temperature since they all have the same temperature $T=|B|/4\pi$. The partition function for this ensemble is given by the path integral sum
\begin{eqnarray}
Z(\beta)&=&\sum_{g\epsilon {\cal S}}\exp (-A^{{\mathrm (e)} }_{\mathrm{LL}}(g))  
\nn \\
&=&\sum_{g\epsilon {\cal S}}\exp \left(-\int_0^\beta  d\tau \int d^3x \sqrt{g_E}L_m[f(r)]\right)
\nn \\
\label{zdef}
\end{eqnarray}
where $m$ denotes the $m-$th Lovelock term. The sum in \eq{zdef} is restricted to the set ${\cal S}$ of all metrics of the form in \eq{fmetric} with the behaviour $[f(a)=0,f'(a)=B]$ and the Euclidean Lagrangian is a functional of $f(r)$. No source term or cosmological constant (which cannot be distinguished from certain form of source) is included since the idea is to obtain a result which depends purely on the geometry. The spatial integration will be restricted to a region bounded by $(D-2)$-spheres $r=a$ and $r=b$, where the choice of $b$ is arbitrary except for the requirement that  within the region of integration the Lorentzian metric must have the proper signature with $t$ being the time coordinate. To evaluate the integral in \eq{zdef}, we first express the \LL lagrangian as functional in terms of $f(r)$. The \LL lagrangian $L_m$ is given by completely anti-symmetrised product of $m$ curvature tensors 
\begin{equation}
L_m=\frac{1}{16\pi} \frac{1}{2^m} \delta^{a_1 b_1 \ldots a_m b_m}_{c_1 d_1 \ldots c_m d_m} R^{c_1 d_1}_{~ a_1 b_1} \cdots R^{c_m d_m}_{~ a_m b_m}
\end{equation} 
For $m=1$, the $L_m$ reduces to $R$, the Einstein-Hilbert lagrangian. For metrics of the type in \eq{fmetric}, the \LL lagrangian has a very nice compact form, and can in fact be written as a total derivative \cite{deser1} (see Appendix \ref{app:LLtotderiv})
\begin{eqnarray}
r^{D-2} L_m &=& \frac{d}{dr} \left[ \frac{(D-2)!}{(D-2m-1)!}(1-f)^m r^{D-2m-1} \right] 
 - \frac{d}{dr}\left[ \frac{(D-2)!}{(D-2m)!}mf^{\prime}(1-f)^{m-1} r^{D-2m} \right]
 \nn \\
 \label{eq:LLtotderiv}
\end{eqnarray}
Thus we have the functional dependence of $L_m$ on $f(r)$. A straight forward calculation shows that
\begin{eqnarray}
- A_{{\mathrm e} m} &=& \frac{\beta \Omega}{16\pi} \int_a^b dr \Biggl\{ \left[\frac{(D-2)!}{(D-2m-1)!}(1-f)^m r^{D-2m-1} \right]^{\prime} 
- \left[ \frac{(D-2)!}{(D-2m)!}mf^{\prime}(1-f)^{m-1} r^{D-2m} \right]^{\prime}   \Biggl \} 
\nonumber \\
&=& \frac{\beta \Omega}{16\pi} \left[\frac{(D-2)!}{(D-2m-1)!} a^{D-2m-1} - \frac{(D-2)!}{(D-2m)!}m B a^{D-2m}  \right] +Q[f(b),f'(b)]
\label{zres}
\end{eqnarray}
where $\Omega$ is the area of a unit $(D-2)$-sphere, $Q$ depends on the behaviour of the metric near $r=b$ and we have used the conditions $[f(a)=0,f'(a)=B]$. The sum in Eq.(\ref{zdef}) now reduces to summing over the values of $[f(b),f'(b)]$ with a suitable (but unknown) measure. This sum, however, will only lead to a factor $Z_Q$ which we can ignore in deciding about the dependence of $Z(\beta)$ on the form of the metric near $r=a$. Using $\beta=4\pi/B$ (and taking $B>0$ for the moment)  the final result can be written in a very suggestive form:
\begin{eqnarray}
Z(\beta) &=& Z_Q \exp \left[ \frac{(D-2)!}{(D-2m)!}\frac{m \Omega}{4} a^{D-2m} - \beta \left( \frac{(D-2)!}{(D-2m-1)!} \frac{\Omega}{16\pi} a^{D-2m-1} \right)  \right] 
\nn \\
\nn \\
&\propto& \exp \left[S^{\mathrm (e)}(a) -\beta E(a)  \right]
\label{zresone}
\end{eqnarray}
with the identifications for the entropy and energy being given by:
\begin{eqnarray}
S^{\mathrm (e)} &=& \frac{(D-2)!}{(D-2m)!}\frac{m \Omega}{4} a^{D-2m} 
\label{actionentropy} \\
E &=& \frac{(D-2)!}{(D-2m-1)!} \frac{\Omega}{16\pi} a^{D-2m-1} \label{actionenergy}
\end{eqnarray}
Comparing the expression of entropy with that of Wald entropy of the black-hole in the corresponding \LL theory \cite{LLenergy},
\begin{eqnarray}
 S_{\mathrm{Wald}} &=& \int_{\cal H} d\Sigma L^{D-2}_{m-1} = \frac{(D-2)!}{(D-2m)!}\frac{m \Omega}{4} a^{D-2m} 
\nn \\ 
 &=& S^{\mathrm (e)}
\end{eqnarray}
which is the result we wanted to demonstrate. In fact, one may even compare $E$ with energy of horizons in \LL theories known in the literature \cite{LLenergy}. The latter is given by
\begin{eqnarray}
E = \int da \int_{\cal H} d\Sigma L^{D-2}_{m} = \frac{(D-2)!}{(D-2m-1)!} \frac{\Omega}{16\pi} a^{D-2m-1}
\nn \\
\end{eqnarray}
which is exactly same as the energy in \eq{actionenergy} obtained above through the analysis of the path integral. For $m=1$ in four dimensions, we recover the known result $E = a/2$ for Einstein's gravity.  Thus, it is very interesting to note that the simple path integral approach, which does not involve the field's equations describing the theory, leads us to the same expressions of Wald entropy and energy of the horizon obtained through other complicated means! Similar result for Wald entropy has been obtained earlier in \cite{Neupane} for specific black hole solutions in Einstein-Gauss-Bonnet gravity and in Riemann squared gravity. 

There are couple of points which need to be stressed regarding this result vis-a-vis previous approaches to computation of entropy.

We begin by noting that there exist several 
different prescriptions in the literature which allow us to calculate black hole entropy in Einstein's theory of gravity (and some of which generalize to higher curvature theories of gravity as well). Examples include: (i) Computation based on hawking radiation of external fields quantized in black hole spacetime by integrating $dS=dM/T(M)$. (ii) Using the relation between the Noether charge associated with the diffeomorphism symmetry of the theory and black hole entropy as proposed by Wald in \cite{Wald}; this is often  taken to be the definition of entropy associated with the horizon in higher curvature theories. (iii) From the Euclidean action of the theory evaluated on-shell for a stationary spacetime with a horizon; this also leads to the expression of Wald entropy \cite{IyerWald}. (We will say more about the relation between this approach and ours towards the end of this section.) (iv) The surface term (analogous to the Hawking-Gibbons  term in GR) in the action,  when evaluated over the horizon surface is equal to the corresponding expression of Wald entropy in the theory \cite{IyerWald}. (v) The entanglement entropy associated with the quantum fields in the background spacetime with a horizon is related to the black hole entropy in GR. (vi) The algebra associated with the Noether charge and diffeomorphisms which leave the horizon geometry intact is known to be same as the Virasoro algebra and hence allows us to calculate the horizon entropy using Cardy formula \cite{carlip}. 

The approach taken in this paper provides us with an \textit{additional} prescription to calculate horizon entropy, introduced and described earlier in \cite{paddy2002}. It has several interesting features not shared by others: (a) We evaluate an off-shell partition function for a class of geometries having a black hole and do \textit{not} use field equations anywhere. This is in contrast with many other prescriptions listed above. (b) We obtain \textit{both} the energy and entropy at one stroke. This is not possible --- as far as we know --- in any of the previous approaches. (c) We do not use the surface term in the action (neither the Gibbons-Hawking term or some equivalent in \LL\ models. [Note that approaches (i),(ii) (v) and (vi) listed in the previous para do not use the surface term either.]

Let us elaborate a little on the last point. Previous work has clearly demonstrated the holographic relationship between the surface and bulk terms in the action, which tells us that the same information is contained in both \cite{holoaction}. So it should \textit{not} come as a surprise that we can get the result without using the surface term.
In fact, it was shown in \cite{paddy2002} that by manipulating the Einstein Hilbert lagrangian $R\sqrt{-g}$, it is possible to express it as a total derivative for a subclass of metrics as given by Eq.(\ref{fmetric}). One can then integrate the lagrangian to obtain the Euclidean action and hence the path integral in the form of Eq.(\ref{zresone}) to read off the energy and entropy of the horizon in Einstein's gravity. We did not need to use the Gibbons-Hawking term in the action in this case.  It is, however, important to see whether the result generalizes to \LL\ models. This is because, 
 in Einstein's gravity, the entropy of a horizon is proportional to its area which has a simple geometrical meaning and therefore very easy to come up with completely different approaches which will lead to $S\propto A$. In such a situation, the robustness of any prescription can then be checked by working in a regime other than Einstein's gravity where the relation between entropy and a geometric property of the horizon is not trivial. We have now shown that the path integral approach considered above is indeed a good prescription to calculate the horizon entropy if one is working within \LL\ theories of gravity. Also, one gets the expression for quasi-local energy as a bonus alongwith the horizon entropy.  In a way, this is nice because the surface term which need to be added to gravitational action --- even in GR --- is \textit{not} unique (see \cite{nelson}), a fact which is not often appreciated in the literature.

Finally, we mention that the scope, procedure and the spirit of the result described above is quite different from some of the previous work using Euclidean action, like e.g., the work by Iyer and Wald \cite{IyerWald}.  We will briefly highlight the differences for placing the current work in proper perspective.

\begin{itemize}
\item
The usual Euclidean approach (like the one in \cite{IyerWald}) essentially evaluates path integral for an action with a surface term in the saddle-point-approximation. (The paper \cite{IyerWald} does not explicitly refer to the path integrals but the only context in which the value of classical action can be interpreted as a thermodynamic variable is in the saddle-point-approximation to the path integral. So this is implicit in the work.)
The \textit{classical} value of the action for a particular \textit{solution} is used to obtain the thermodynamic interpretation. In contrast, we do \textit{not} use the saddle point approximation or assume validity of classical field equations anywhere; therefore our result captures at least certain aspects of the  \textit{off-shell} physics. Of course, since the exact path integral is intractable we also need to make some approximation; we evaluate the sum over a subset of geometries which possess a horizon. This is quite different mathematically and conceptually from approximating the path integral by a saddle point value since we do not have to assume the validity of  field equations or a classical solution for our evaluation. 
\item
For the same reason, we use the expression for the off-shell Noether charge; that is, we do not assume equations of motion to set  part of this Noether charge to zero, unlike in the approach taken by \cite{IyerWald}. The Noether charge we use is identically conserved off shell (due to diffeomorphism invariance of the lagrangian) without our assuming the validity of equations of motion \cite{paddyaspects}. 
\item
We do not add any surface term (analogous to $K\sqrt{h}$ in the case of general relativity) to the \LL\ action while the previous Euclidean approaches depended on the addition of the surface term. This is important because the geometric nature and uniqueness of surface term for \LL\ models is unclear. Even in GR, the surface term is not unique. The main purpose of the surface term is to cancel off the time derivatives of the metric variations at the spacetime boundary arising from the variation of the action. There are several surface terms in GR other than $K\sqrt{h}$ which satisfy this criteria. Although $K\sqrt{h}$ turns out to be the simplest among them, it is not well understood why it should be equal to entropy when evaluated on the horizon. It is interesting that we do not have to use any surface term.
\item
We were able to evaluate the sum over the restricted class of metrics (without resorting to saddle point approximation)  only because the lagrangian turns out to be a total derivative for the class of metrics considered. \textit{This, by itself, is a deep result and could not have been guessed a priori for the \LL\ models.} Performing the path integral, we find contributions to the action \textit{both} from a quasi-local energy term and a entropy term which allows us to clearly identify both the energy and entropy of the system separately and the partition function appears as the exponential of the free energy. The previous approaches, instead, obtained only the entropy of the system. 
\end{itemize}
These aspects make our analysis complementary to the approaches based on the saddle-point-approximation.

\section{Conclusions}

We have analyzed two aspects of black hole entropy for higher curvature theories of gravity, one from the point of view of matter entropy, and other in terms of a Euclidean path integral defining a partition function. These issues have already been studied in Einstein theory described by the conventional Einstein-Hilbert action, and our aim in this note was to generalize  the analyses for \LL\ models (and $F(R)$ gravity) and see how the entropy so obtained matches with Wald entropy, which is generally accepted as the correct definition of entropy for bifurcate Killing horizons in arbitrary theories of gravity.

As explained in the introduction,
our motivation goes beyond a mere generalization of certain known results to modified actions and higher dimensions.
(In fact,  we do not presume that the class of modified actions we have considered would have any practical relevance in realistic situations.) Rather, the aim here is to use these more general results to better understand the origin of horizon entropy in Einstein theory itself.
We are somewhat handicapped in discriminating between different approaches to calculate the entropy within the limited context of Einstein's theory because an entropy proportional to area can easily arise in completely different contexts. For example, it is known that entanglement entropy associated with a bounded region of space would always scale as area of the boundary of that region, and similarly area scaling also arises independently of field equations, from purely kinematic considerations, when one analyses the phase space available to ordinary thermal systems near a horizon \cite{areaentropysanved}.  
The only acid test for the validity of any particular approach for calculating horizon entropy is that it should reproduce the Wald entropy of, say, at least \LL models which are natural generalizations of Einstein's theory. It is with this motivation that we studied the two different approaches in this paper.

We found that if one evaluates the entropy of matter which is on the verge of collapsing to a black hole, then this matter entropy does not necessarily match with Wald entropy. For  a \LL model of order $m$,   the entropy is proportional to Wald entropy with the proportionality constant depending on order $m$ of the \LL polynomial. 
Hence when we add up \LL actions of different orders, the overall proportionality breaks down.
(However, the entropy matched with Wald entropy for the $F(R)$ models.) In the second part of this paper, we considered another approach which attempts to look at an ``intrinsic" approach to horizon entropy, by showing that the Euclidean action for \LL models, in spherical symmetry, becomes a total derivative and can be written as an expression for free-energy from which entropy $S$ and energy $E$ can be read-off. The expression for $S$ then matches with Wald entropy, and $E$ reduces to the quasi-local expressions known for special cases in the literature.

The main implication of this note is that matter entropy of a system is not necessarily the only contribution to entropy of a black hole formed from collapse of that system. Although there exists in the literature several entropy bounds on matter entropy, our model calculation gives precise expressions for matter entropy and its relation to Wald entropy, rather than just an inequality. Besides that, our analysis also highlights a possible connection between horizon entropy and the ``equation-of-state" $P(T)$ satisfied by the degrees of freedom residing on the horizon surface. This relation is worth investigating further. Moroever, as was indicated in the text, the pressure which we obtain in the horizon limit is precisely the pressure encountered in the membrane paradigm. In fact, in a subsequent publication, we shall present the full set of transport coefficients determining the dynamics of a perturbed horizons, as is done in the membrane paradigm.

%%%%%%%%%%%%%%%%%%%%%%%%%%%%%%%%%%%%%%%%%%%%%%%%%%%%%%%%%%%%%%%%%%%%%%%%%%%%%%%%%%%%%%%%%%%%%
\section*{Acknowledgements}
%%%%%%%%%%%%%%%%%%%%%%%%%%%%%%%%%%%%%%%%%%%%%%%%%%%%%%%%%%%%%%%%%%%%%%%%%%%%%%%%%%%%%%%%%%%%%
The authors thank Olivera Miskovic and Rodrigo Olea for useful clarifications regarding the surface stress tensor in \LL gravity. SK is supported by a Fellowship from the Council of Scientific and Industrial Research (CSIR), India. The research of DK is funded by NSERC of Canada, and Atlantic Association for Research in the Mathematical Sciences (AARMS). The research of TP is partially supported by the J.C. Bose fellowship of DST, India

\vspace{0.2cm}

\appendix
\section*{Appendix:}
\vspace{0.2cm}
%%%%%%%%%%%%%%%%%%%%%%%%%%%%%%%%%%%%%%%%%%%%%%%%%%%%%%%%%%%%%%%%%%%%%%%%%%%%%%%%%%%%%%%%%%%%%%%%%%%%%%%%
\section{$F(R)$ gravity} \label{app:fR}
%%%%%%%%%%%%%%%%%%%%%%%%%%%%%%%%%%%%%%%%%%%%%%%%%%%%%%%%%%%%%%%%%%%%%%%%%%%%%%%%%%%%%%%%%%%%%%%%%%%%%%%%
In the main text of this paper, we calculated the matter entropy of gravitating shells in \LL models of gravity. However, the same procedure involved can be used to determine the matter entropy in $F(R)$ theories of gravity which we will briefly discuss below.

The $F(R)$ theories of gravity are characterized by the lagrangian of the form $L = F(R)$ where $F(R)$ is a well behaved (infinitely differentiable) function of Ricciscalar $R$. Spherically symmetric black hole solutions in such theories have been studied in the literature, but we shall not need detailed form of these solutions for the analysis. The Wald formula for black hole entropy in such theories is given by 
\begin{equation}
S_{\mathrm{Wald}} = \frac{F^{\prime}(R(r_H))}{4} A_H
\end{equation}
where the $F^{\prime}=\DM F/\DM R$ and $A_H$ is the $(D-2)$ dimensional area of the black hole horizon. We wish to compare this expression for Wald entropy with the entropy of the system on the verge of forming a black hole, as described in the previous section.

The junction conditions in F(R) gravity are given by \cite{junction-fr}
\begin{equation}
8 \pi s_{\mu \nu} = 
\Biggl \langle 
F^{\prime}(R) \l( K h_{\mu \nu} - K_{\mu \nu} \r) + h_{\mu \nu} F^{\prime \prime}(R) n^k \partial_k R
\Biggl \rangle_i
\nn
\end{equation}
which is to be supplemented by an additional condition of continuity of Ricciscalar across a given surface. In fact, if one assumes that the function $F(R)$ is well behaved, then one can immediately write down the final result for entropy of this configuration, simply by comparing it with the Einstein case discussed in detail above. This immediately yields the result that, in the near horizon limit, 
\begin{eqnarray}
S_{\mathrm{matter}} = S_{\mathrm{Wald}}
\end{eqnarray}
and hence the correspondence between matter and Wald entropy holds for $F(R)$ theories. Although the extrinsic approach lead to the exact expression of Wald entropy for the matter entropy, in general the intrinsic approach considered in the second half of the paper will not hold for $F(R)$ theories. This is because although $R$ can be expressed as a total divergence for metrics of the form \eq{fmetric}, any arbitrary function of $R$ will not be a total divergence. This also highlights the special status of \LL theories in the sense that results in Einstein gravity easily generalize to \LL theories whereas in the case of $F(R)$ theories they do not.
%%%%%%%%%%%%%%%%%%%%%%%%%%%%%%%%%%%%%%%%%%%%%%%%%%%%%%%%%%%%%%%%%%%%%%%%%%%%%%%%%%%%%%%%%%%%%%%%%%%%%%%%
\section{Einstein-Gauss-Bonnet (EGB) gravity}
%%%%%%%%%%%%%%%%%%%%%%%%%%%%%%%%%%%%%%%%%%%%%%%%%%%%%%%%%%%%%%%%%%%%%%%%%%%%%%%%%%%%%%%%%%%%%%%%%%%%%%%%
In this appendix, we will show the explicit calculation of matter entropy in the case of $(m=2)$ Einstein-Gauss-Bonnet gravity (for known black hole solutions, see \cite{deser2}). The surface stress tensor in EGB theory is given by \cite{junctionGB}
\begin{eqnarray}
8 \pi t_{\mu \nu} &=& 8 \pi t^{(E)}_{\mu \nu} - 2 \alpha \Biggl \langle  J h_{\mu \nu} - 3 J_{\mu \nu}  + 2 \hat P_{\mu \rho \sigma \nu} K^{\rho \sigma} \Biggl \rangle
\nn \\
J_{\mu \nu} &=& \frac{1}{3} \Biggl[ 2 K K_{\mu \rho} K^{\rho}_{\nu} + K_{\mu \nu} \l( \mathrm{tr} K^2 - (\mathrm{tr} K)^2 \r) - 2 K_{\mu \rho} K^{\rho \sigma} K_{\sigma \nu} \Biggl]
\nn \\
\hat P_{\mu \rho \sigma \nu} &=& \hat R_{\mu \rho \sigma \nu} + 2 \hat R_{\rho [\sigma} h_{\nu] \mu} - 2 \hat R_{\mu [\sigma} h_{\nu] \rho} + \hat R h_{\mu [\sigma} h_{\nu] \rho}
\label{junctionconditionGB}
\end{eqnarray}
where $\hat R_{\mu \rho \sigma \nu}$ etc. indicates that these quantities are to be evaluated for the induced metric $h_{\mu \nu}$. After a lengthy but straightforward computation, one can show that
\begin{eqnarray}
3 J_{\mu \nu} - J h_{\mu \nu} &=& j_1 u_{\mu} u_{\nu} + j_2 q_{\mu \nu}
\end{eqnarray}
where
\begin{eqnarray}
j_1 &=& - \frac{(D-3)(D-4)}{3 (D-2)^2} K_{\perp}^3 
\nn \\
j_2 &=& \Biggl[ -K_{\hat 0 \hat 0} + \frac{(D-5)}{3 (D-2)} K_{\perp} \Biggl] \frac{(D-3) (D-4) K_{\perp}^2}{(D-2)^2} 
\nn \\
\end{eqnarray}
and $K_{\perp} = q^{\mu \nu} K_{\mu \nu} = (D-2) \sqrt{b}/r$. We therefore note that both $j_1$ and $j_2$ are finite (in fact vanishing) at $f=0=b$. Hence these will not contribute to entropy. Indeed, the only relevant contribution comes from the $\hat P_{\mu \rho \sigma \nu} K^{\rho \sigma}$ term in $t_{\mu \nu}$. Using the fact that $r=$const. surfaces are $\mathbb{R} \times \mathbb{S}^{D-2}$, this term evaluates to:
\begin{eqnarray}
\hat P_{\mu \rho \sigma \nu} K^{\rho \sigma} &=& \frac{(D-3)(D-4)}{2 r^2} \Biggl[ 
K_\perp u_\mu u_\nu +
\l( K_{\hat 0 \hat 0} - \frac{D-5}{D-2} K_{\perp} \r) q_{\mu \nu}
\Biggl]
\end{eqnarray}

We now have all the quantities needed to evaluate the divergent contributions to $t_{\mu \nu}$. By inspection of above expressions, we see that divergences go as $1/\sqrt{f}$ (despite the GB contribution being cubic in extrinsic curvature), and the divergent pieces can be extracted out as:
\begin{eqnarray}
8 \pi [\rho_n]_{\mathrm{div}} =  8 \pi \l[ t_{\hat 0 \hat 0} \r]_{\mathrm{div}} &=& 8 \pi \l[ t_{\mu \nu} u^{\mu} u^{\nu} \r]_{\mathrm{div}} = 0 
\nn \\
\nn \\
8 \pi [P_n]_{\mathrm{div}}\; \delta_B^a = 8 \pi \l[ t^{\ \ A}_{(n) B} \r]_{\mathrm{div}} &=& - K_{\hat 0 \hat 0} \delta^A_B - [4 \alpha \hat P^{A \rho}_{\ \ \ \sigma B} K^{\rho}_{ \sigma}]_{\mathrm{div}}
\nn \\
&=& - K_{\hat 0 \hat 0} \l[ 1 + \frac{2 \alpha (D-3) (D-4)}{r^2} \r] \delta^A_B
\label{pressureGB}
\end{eqnarray}
Hence, the only non-zero contribution to the matter entropy comes only from the pressure term as in the case of Einstein gravity (see \eq{eq:sabeinsdiv}). No further computation needs to be done, since we can simply obtain the entropy by comparing the above expressions with Eqs.~(\ref{eq:sabeinsdiv}) in Einstein theory. This immediately allows us to ``read-off" the matter entropy as
\begin{eqnarray}
S_{\mathrm{matter}} = \frac{A_H}{4} \l[ 1 + \frac{2 \alpha (D-3) (D-4)}{r^2} \r] \label{matterentropyGB}
\end{eqnarray}
On the other hand, the Wald entropy for EGB theory can be shown to give
\begin{eqnarray}
S_{\mathrm{Wald}} &=& \frac{A_H}{4} \l[ 1 + \frac{2 \alpha (D-3) (D-2)}{r^2} \r]
\nn \\
&=& S_{\mathrm{matter}} + \frac{(D-3) \alpha A_H}{r^2} \label{waldentropyGB}
\end{eqnarray}
We see that $S_{\mathrm{Wald}} > S_{\mathrm{matter}}$ (assuming $\alpha>0$). 

%%%%%%%%%%%%%%%%%%%%%%%%%%%%%%%%%%%%%%%%%%%%%%%%%%%%%%%%%%%%%%%%%%%%%%%%%%%%%%%%%%%%%%%%%%%%%%%%%%%%%%%%
\section{\LL lagrangian as a total derivative for spherically symmetric spacetimes} \label{app:LLtotderiv}
%%%%%%%%%%%%%%%%%%%%%%%%%%%%%%%%%%%%%%%%%%%%%%%%%%%%%%%%%%%%%%%%%%%%%%%%%%%%%%%%%%%%%%%%%%%%%%%%%%%%%%%%

Consider a $D$ dimensional spacetime $M^D$, which is a direct product of a two dimensional manifold $M^2$ with a $(D-2)$ maximally symmetric sub-manifold $\mathcal{K}^{D-2}$ with sectional curvature $k=\pm1,0$ and coordinates $z^A$. The metric can be written as
\begin{equation}
 ds^2 = g_{ab}(y)dy^a dy^b + r^2(y)q_{AB}dz^A dz^B
\label{maxsymmetric}
\end{equation}
where $a, b = 0, 1$; $i, j = 2, ..., D-1$ and $r(y)$ is a scalar on $M^2$. Then for this metric the \LL lagrangian of an arbitrary order $m$ can be written in a very compact form \cite{maeda}
\begin{equation}
 L_m = \frac{(D-2)!}{(D-2m)!} \Psi_m
\end{equation}
where 
\begin{eqnarray}
 \Psi_m &=& (D-2m)(D-2m-1) \psi^m - 2(D-2m)m\left[ \frac{D^2r}{r}\right] \psi^{m-1} 
 \nn \\
 &+&  2m(m-1)\left[ \frac{(D^2r)^2 - (D^aD_br)(D^bD_ar)}{r^2} \right] \psi^{m-2} 
+ m \; {}^2R \; \psi^{m-1}
\nn \\
\nn \\
\psi &=& \frac{k - (Dr)^2}{r^2}
\end{eqnarray}
Here $D_a$ is a metric compatible linear connection on $M^2$, $(Dr)^2 := g^{ab}(D_ar)(D_br)$, $D^2r := D^aD_ar$ and $^2R$ is the Ricci scalar on $M^2$. We can now use the above expression with metric \eq{fmetric} to express the \LL action in a compact form. We have, $(Dr)^2=f(r)$, $D^2r=f^{\prime}$, $(D^aD_br)(D^bD_ar) = f'^{2}/2$ and ${}^2R = -f^{\prime \prime}$, which leads to
\begin{eqnarray}
 \Psi_m &=& (D-2m)(D-2m-1) \psi^m - 2(D-2m)m\left[ \frac{f^{\prime}}{r}\right] \psi ^{m-1} 
+ m(m-1)\left[ \frac{f'^2}{r^2} \right] \psi^{m-2} - mf^{\prime \prime} \psi^{m-1}
\nn \\
\psi &=& \frac{1-f}{r^2}
\end{eqnarray}
%\end{widetext}
Further, using (with $q=1-f$)
\begin{eqnarray}
- \frac{d}{dr} \left\{ m f^{\prime} q^{m-1} r^{D-2m} \right\} = 
&-& m f^{\prime \prime} q^{m-1} r^{D-2m} 
+ m(m-1) f'^2 q^{m-2} r^{D-2m} 
\nn \\
&-& m(D-2m) f^{\prime} q^{m-1} r^{D-2m-1}
\nn
\end{eqnarray}
and
\begin{eqnarray}
\frac{d}{dr}\left\{ (D-2m) q^{m} r^{D-2m-1} \right\} &=&  (D-2m)(D-2m-1) q^{m} r^{D-2m-2} 
\nn \\
&-& m(D-2m) f^{\prime} q^{m-1} r^{D-2m-1}
\end{eqnarray}
we can easily write the \LL lagrangian as a total derivative as given in \eq{eq:LLtotderiv}.

%%%%%%%%%%%%%%%%%%%%%%%%%%%%%%%%%%%%%%%%%%%%%%%%%%%%%%%%%%%%%%%%%%%%%%%%%%%%%%%%%%%%%%%%%%%%%%%%%%%%%%%%

%%%%%%%%%%%%%%%%%%%%%%%%%%%%%%%%%%%%%%%%%%%%%%%%%%%%%%%%%%%%%%%%%%%%%%%%%%%%%%%%%%%%%%%%%%

\end{document}